\documentclass{article}

\usepackage[english]{babel}

\usepackage[letterpaper,top=2cm,bottom=2cm,left=3cm,right=3cm,marginparwidth=1.75cm]{geometry}

\usepackage{amsmath}
\usepackage{amsfonts}
\usepackage{graphicx}
\usepackage{hyperref}
\usepackage{booktabs}
\usepackage{bbding}
\usepackage{appendix}
\usepackage{multirow}
\usepackage{caption}
\usepackage{subcaption}
\usepackage{svg}
\usepackage{dirtytalk}
\usepackage[numbers]{natbib}
\usepackage[nohyperlinks, printonlyused, nolist]{acronym}
\usepackage{authblk}
\hypersetup{colorlinks=true, allcolors=blue}

\title{Reconciling AI Performance and Data Reconstruction Resilience for Medical Imaging}
\author[1,$\star$]{Alexander Ziller}
\author[1]{Tamara T. Mueller}
\author[1]{Simon Stieger}
\author[1]{Leonhard Feiner}
\author[1]{Johannes Brandt}
\author[1]{Rickmer Braren}
\author[1,2]{Daniel Rueckert}
\author[1,2,3]{Georgios Kaissis} 
\affil[$\star$]{alex.ziller@tum.de}
\affil[1]{Technical University of Munich}
\affil[2]{Imperial College London}
\affil[3]{Helmholtz Munich}
\begin{document}

\begin{acronym}
    \acro{ai}[AI]{Artificial Intelligence}
    \acro{dp}[DP]{Differential Privacy}
    \acro{gdpr}[GDPR]{General Data Protection Regulation}
    \acro{mia}[MIA]{Membership Inference Attack}
    \acro{sota}[SOTA]{state-of-the-art}
    \acro{fpr}[FPR]{False Positive Rate}
    \acro{tpr}[TPR]{True Positive Rate}
    \acro{fnr}[FNR]{False Negative Rate}
    \acro{auc}[AUC]{area under the receiver-operator curve}
    \acro{mcc}[MCC]{Matthews' Correlation Coefficient}
    \acro{roc}[ROC]{receiver-operator characteristic}
    \acro{mri}[MRI]{magnetic resonance imaging}
    \acro{ct}[CT]{computed tomography}
    \acro{rero}[ReRo]{reconstruction robustness}
    \acro{ssim}[SSIM]{structural similarity}
    \acro{pet}[PET]{privacy-enhancing technology}
    \acroplural{pet}[PETs]{privacy-enhancing technologies}
    \acro{dra}[DRA]{data reconstruction attack}
    \acro{dpsgd}[DP-SGD]{Differentially Private Stochastic Gradient Descent}
    \end{acronym}

\date{}

\maketitle
% \begin{figure}[h!]
%     \centering
%     \includegraphics[width=0.5\textwidth]{figures/headerfigure.pdf}
%     \caption{
%     % \textbf{Reconstructions extracted from the gradients in an AI training process.} On the top row the original images of political figures around the globe. In the middle reconstructions of the training data from a gradient in an AI training process. Gradients are available to the server of a federated learning process. On the bottom we see that these reconstructions can be completely impeded by the use of Differential Privacy even at vast privacy budgets. Here, we used $\varepsilon=10^{9}$ at $\delta=0.01$.
%     This figure demonstrates the dangers of training models without any privacy protection. The original images used to train the model (top row) can be perfectly reconstructed by a realistic adversary (middle row). Introducing even a very modest amount of privacy protection to model training renders these attacks ineffective.
%     }
%     % \label{fig:celeba}
% \end{figure}
%TC:ignore
\begin{abstract}
Artificial Intelligence (AI) models are vulnerable to information leakage of their training data, which can be highly sensitive, for example in medical imaging. Privacy Enhancing Technologies (PETs), such as Differential Privacy (DP), aim to circumvent these susceptibilities. DP is the strongest possible protection for training models while bounding the risks of inferring the inclusion of training samples or reconstructing the original data. DP achieves this by setting a quantifiable privacy budget. Although a lower budget decreases the risk of information leakage, it typically also reduces the performance of such models. This imposes a trade-off between robust performance and stringent privacy. Additionally, the interpretation of a privacy budget remains abstract and challenging to contextualize. In this study, we contrast the performance of AI models at various privacy budgets against both, theoretical risk bounds and empirical success of reconstruction attacks. We show that using very large privacy budgets can render reconstruction attacks impossible, while drops in performance are negligible. We thus conclude that not using DP --at all-- is negligent when applying AI models to sensitive data. We deem those results to lie a foundation for further debates on striking a balance between privacy risks and model performance.
\end{abstract}
%TC:endignore

\section{Main}
The rapid rise of \ac{ai} applications in medicine promises to transform healthcare, offering improvements ranging from specific applications, such as more precise pathology detection, outcome prediction, to the promise of general medical \ac{ai} \cite{laang2023artificial,wang2023deep,al2023machine,singhal2023large,jiang2023health}. 
However, recent results highlight a significant vulnerability: \ac{ai} models may disclose details of their training data. 
This can happen either inadvertently or be forced through attacks by malicious third parties, also called adversaries. 
Among the most critical attacks are \textit{data reconstruction attacks}, where the adversary attempts to extract training data from the model or its gradients \cite{geiping2020inverting,yin2021see,fowl2021robbing,boenisch2023curious,wang2021variational,haim2022reconstructing,carlini2023extracting,buzaglo2023deconstructing}.
Such attacks harbour distinct risks. 
On one hand, a successful data reconstruction attack severely undermines the trust of patients whose data is exposed.
This not only jeopardises the relationship between medical practitioners and patients, but likely also diminishes the willingness of patients to make their health data for the training of \ac{ai} models or for other research purposes available. 
This is problematic since the success of \ac{ai} models in medicine is dependent on the availability of large and diverse real-world patient datasets.
On the other hand, a successful attack can also constitute a breach of patient data privacy regulations.

%Despite regulations pertaining to general (non-medical) data varying considerably between legislations
While privacy laws vary globally, the protection of health data is generally considered of high importance. % in most countries. 
For example, the European Union's \ac{gdpr} declares the protection of \textit{personal data} as a fundamental right.
Notably, these laws deem the removal of personal identifiers (e.g. name or date of birth) --\textit{de-identification}-- sufficient protection. 
However, it has been demonstrated on several occasions that commonly used de-identification techniques such as anonymisation, pseudonymisation, or $k$-anonymity are vulnerable to re-identification attacks \cite{narayanan2008robust, cohen2020towards, cohen2022attacks}.
This also holds true in the case of medical imaging data.
For example, the facial contours of a patient can be obtained from a reconstructed \ac{mri} scan even if their name has been removed from the record, thus enabling their re-identification from publicly available photographs \cite{schwarz2019identification}. 
Figuratively, this is analogous to considering passport photos without additional information not as personal data.
Arguably, this highlights the tension between what is considered \say{private} in a legal sense and what individuals consider acceptable in terms of informational self-determination. 
We thus contend that AI systems which process sensitive data should not only rely on de-identification techniques but also implement \acp{pet}, i.e. technologies which furnish an objective or formal guarantee of privacy protection.

\begin{figure}[t!]
    \centering
    \includegraphics[width=0.95\textwidth]{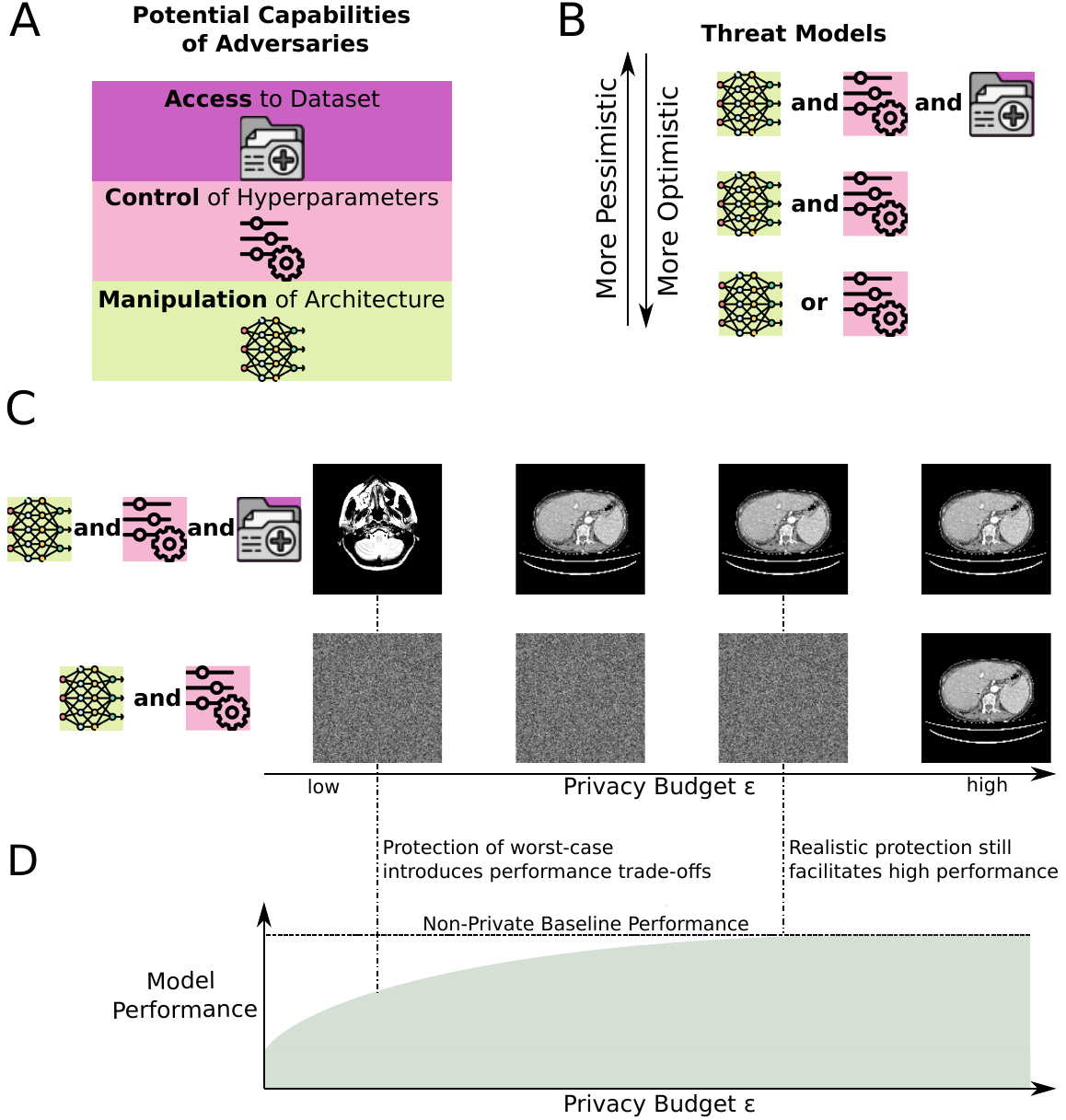}
    \caption{\textbf{Comparison of a worst-case and a realistic threat model.} \textbf{(A)} Adversaries can have various capabilities depending on the setting \textbf{(B)} The combination of the adversary's capabilities defines the threat model. In a worst-case analysis, they have all capabilities. However, access to the database is a pessimistic, practically irrelevant scenario. \textbf{(C)} The necessary privacy protection depends on the threat model. In a worst-case threat model, the adversary only needs to match the model and gradient to an image in the database. In a practically more relevant scenario, the image must be reconstructed from the model and gradient. Here, much less privacy protection is necessary \textbf{(D)} The more stringent the privacy protection is chosen, the higher the impacts on the model performance are. Thus, if a realistic threat model is considered appropriate, models can perform better. }
    \label{fig:overview}
\end{figure}

% \subsection{Necessity of Limiting the Privacy Risk in AI Training}
\subsection{Differential Privacy as the Strongest Possible Privacy Preservation}
Among the \aclp{pet}, \ac{dp} \cite{dwork2014algorithmic} is considered the optimal protection for training \ac{ai} models while moderating the privacy risk faced by participating patients.
This is because it has several appealing properties: \ac{dp} is the only technique providing mathematical guarantees for an upper bound of the success of reconstructing data samples \cite{balle2022reconstructing,kaissis2023bounding}. Also, it satisfies requirements imposed by regulations like the \ac{gdpr} concerning re-identification \cite{cohen2020towards,nissim2021privacy}.
Moreover, the privacy guarantees of \ac{dp} systems cannot be degraded through the use of side information or through post-processing by additional computation (two notable vulnerabilities of traditional de-identification schemes).
Last but not least, \ac{dp} satisfies \textit{composability}, i.e., the notion that privacy guarantees degrade predictably when multiple \ac{dp} algorithms are executed on the same dataset. 
This enables the concept of a \say{privacy budget}, which makes the cumulative re-identification risk quantifiable and which can be set depending on policy or preference.
We note that this ability to moderate risks stemming from \ac{ai} applications is particularly beneficial, as this ability is also mandated by recent legal frameworks such as the European AI act \cite{eu2021aiact}.
These properties are leading to its increasing adoption in industry and government applications \cite{foote2019releasing,aktay2020google}.

Despite these benefits, the effective and efficient implementation of \ac{dp} in large-scale \ac{ai} systems also presents a series of challenges.
\ac{dp} has been criticised for the fact that the choice of an appropriate privacy budget is delicate. 
Higher privacy budgets correspond to lower levels of privacy protection and thus increase the risk of successful attacks, while lower privacy budgets limit the information available for training. 
By that, it introduces a new challenge, namely a trade-off between privacy protection and model performance, i.e. diagnostic accuracy for a given use-case. Furthermore, this trade-off also depends on the specific input data and learning task, which can vary significantly between scenarios.
Arguably, concerns about reduced model performance are a probable reason why, despite its many benefits, \ac{dp} is not yet widely implemented in medical AI workflows. 
After all, finding  a trade-off between diagnostic accuracy and privacy represents a complex technical and ethical dilemma. 
%A balance must be found between protecting the privacy of the patients who contributed their data and the model's performance on future patients.
This dilemma is also pictured at its extreme as \ac{dp} is underlaid by a worst-case set of assumptions. 
These assumptions, also called a \textit{threat model}, include an adversary who is able to deeply manipulate and interfere with the dataset, the training process, model architecture, and (hyper-)parameters, and has access to all parameters of the \ac{dp} algorithm (mechanism).
Moreover, the canonical \ac{dp} adversary is not assumed to execute a data reconstruction attack but a much simpler type of attack, namely a \textit{membership inference attack} which attempts to determine whether a specific individual's data (which is available to the adversary) was included in the training dataset or not.
Since there are only two possible outcomes of such an attack (member or non-member), membership inference must only reveal a single \textit{bit} of information, compared to a data reconstruction attack, which must successfully reveal a much larger record (e.g. an image comprising several million bits of information).
Evidently, although worst-case assumptions are prudent for the theoretical modelling of adversaries, the \ac{dp} threat model is unlikely to ever be encountered in practice. 
Moreover, the aforementioned membership inference attack in which the adversary has access to a target record and tries to determine whether it was used for training a specific model is arguably of very low practical relevance.
Instead, data reconstruction attacks are likely perceived as a significantly more relevant privacy threat by patients. 
Moreover, realistic adversaries in the medical setting (where data is strongly guarded) can likely be assumed to not have access to the training data (as they would have little incentive to attack a model otherwise).

In this paper, we investigate the question of whether the aforementioned typical \ac{dp} threat model might be too pessimistic for practical use-cases and, by that, impose unnecessary privacy/performance trade-offs.
To investigate this hypothesis, our work studies the privacy/performance characteristics of AI models trained on large-scale medical imaging datasets under more realistic threat models which still allow for strong privacy protection but represent a \say{step down} from the worst-case assumptions made by \ac{dp}.
Our main finding is that, even in complex medical imaging tasks, it is possible to train \ac{ai} models with excellent diagnostic performance while still defending against data reconstruction attacks and thus patient re-identification.
We achieve this by training models under privacy budgets which would be considered too large to offer any protection against the threats considered under the worst-case \ac{dp} threat model.
This provides evidence that there is a recommendation to by default train \ac{ai} models with \ac{dp} protection.
Therefore, although more restrictive privacy budgets than the ones used in our study remain relevant for use-cases in which protection against membership inference is explicitly required, our findings highlight the existence of an additional option:
When high model performance is required but cannot be achieved without relinquishing membership inference protection, our findings offer a compromise whereby an important and relevant class of attacks can be defended against while fulfilling the requirement for high diagnostic accuracy.

%PUT IN CONCLUSION
%We are hopeful that our findings provide the basis for an informed discourse between ethicists, patients etc. ...

\subsection{Risk Quantification of AI Models}

\begin{table}[t]
    \centering
    \begin{tabular}{llll}\toprule
         & Worst-case & Relaxed & Realistic \\ \midrule
        Model Architecture \& Weight & yes & yes & yes  \\
        % Weight Manipulation & yes & yes & yes \\
        Hyperparameter & yes & yes & yes \\
        Dataset Access & yes & partially & no  \\
        Perfect Reconstruction Algorithm & n.n. & yes & no  \\
        \midrule
        Risk Analysis & Theoretical & Theoretical & Empirical \\
        \bottomrule
    \end{tabular}
    \caption{Overview of the capabilities of an adversary in the threat models analysed in this study.}
    \label{tab:threat_models}
\end{table}
As alluded to in the previous section, \ac{dp} allows for a quantifiable reduction in the risk of privacy attacks associated with the training of \ac{ai} models. 
Before introducing our main results, we provide a more concrete explanation of how the risks of re-identification through a data reconstruction attack are moderated by \ac{dp}.
Concretely, we avail ourselves of the framework of \ac{rero}, which will allow us to formulate an upper bound on the success rate of data reconstruction attacks against \ac{ai} models trained with \ac{dp} under the specific threat models discussed below.

\ac{rero} was introduced by \cite{balle2022reconstructing}. 
It is a guarantee pertaining to an algorithm which processes sensitive data, e.g. an \ac{ai} model trained with \ac{dp}.
Intuitively, if at most a proportion $0\geq \gamma \leq 1$ of the total samples used to train the model can be successfully reconstructed by an adversary with a reconstruction error lower than $\eta \geq 0$, then the model satisfies $(\eta,\gamma)$-\ac{rero}. 
Recent works have proven that all models trained with \ac{dp} automatically satisfy \ac{rero} and that for certain settings, it is possible to directly quantify the upper bound for $\gamma$ \cite{balle2022reconstructing,hayes2023bounding,kaissis2023bounding}.
In other words, \ac{dp} automatically provides strong and quantifiable protection against data reconstruction attacks.

In the results section below, we will study the \ac{rero} guarantees of models trained with \ac{dp} under three distinct sets of assumptions about the capabilities of the adversary, i.e. three distinct threat models:
\begin{enumerate}
    \item The \textbf{worst-case} threat model: This corresponds to the adversary usually considered in \ac{dp}, namely one who has unbounded computational abilities, can deeply manipulate the model's (hyper-)parameters and has access to the target image itself, which they can use to attack the model. 
    Evidently, this threat model is not realistic (as an adversary who has access to the target point would not need to attack the model), but is used to provide guarantees when \say{all bets are off}, i.e. in a so-called \textit{privacy auditing} scenario when one is interested in the absolute worst-case behaviour of a system.
    \item The \textbf{relaxed} threat model \cite{kaissis2023optimal}: This threat model is still quite pessimistic, as it still assumes unbounded computational ability and access to model (hyper-)parameters.
    However, this adversary only has restricted access to the dataset, notably, they cannot use the target image itself to attack the model.
    Although it renders this threat model more appropriate for scenarios where the dataset can be safely assumed to be kept secure, e.g. in a hospital's database, it still makes assumptions, which are not encountered in any practical scenario. 
Most importantly, the adversary has a black-box reconstruction algorithm, which yields either a perfect reconstruction or fails, and the only decision the adversary has to make is whether the reconstruction was indeed the target data.
%We note that although this threat model is already less restrictive, it is still extremely pessimistic, as all images but the target image would be available to the attacker. 
    The term \textit{relaxed} stems from security research, where a \textit{relaxation} denotes a weakening of a security assumption.
    \item The \textbf{realistic} threat model: The final threat model considers an adversary with unbounded computational ability and the power to manipulate model (hyper-)parameters but only very limited access to information about the dataset.
    For example, the adversary can know the dimensions of the images to be reconstructed but not any of their contents.
    We note that even this threat model is relatively pessimistic, as it assumes an active adversary who is trusted and therefore the actions are not reviewed by other participants. 
Such adversaries could manipulate the model to their advantage in order to reconstruct training data.
    In many cases, this could be detected simply by inspecting the model architecture.
    Nonetheless, we use this threat model as it is conceivable that such adversaries can exist in, e.g. federated learning settings in untrustworthy consortia. 
\end{enumerate}
Of note, it is possible to provide theoretical bounds on the reconstruction attack success rate in both the worst-case and the relaxed threat models using the techniques presented in \cite{hayes2023bounding, kaissis2023bounding}.
For the realistic threat model, we assess the attack success rate empirically.
A concise overview of the aforementioned threat models is provided in Table \ref{tab:threat_models}.

In summary, while conservative (i.e. worst-case or relaxed) threat models are important tools in security research because they allow one to derive closed-form bounds on the attack success rate of very powerful adversaries, such threat models are in most reasonable scenarios too pessimistic. 
The key contribution of our work is to investigate the realistic risks posed by a type of adversary who is still very powerful but can be reasonably assumed to exist in real-world medical \ac{ai} model training use cases. 
In the next section, we will show that perfectly defending against such adversaries is possible while maintaining a competitive diagnostic model performance with that of a model trained without any privacy protection.

\section{Results}
\subsection{Setup}
\subsubsection{Datasets}\label{sec:datasets}
We begin by introducing our rationale for choosing specific datasets for our experiments. 
We identified four characteristics of medical imaging datasets, which reoccur frequently: (1) Datasets are often \textbf{small} compared to non-medical datasets. For example, most medical \ac{ai} algorithms, which are currently approved by the US Food and Drug Administration (FDA), are trained on less than $1\,000$ data samples \cite{fdawebpage}. (2) Diagnoses occur with very different frequencies, leading to often imbalanced datasets skewed toward more common diagnoses. In segmentation tasks, this may happen due to different spatial extensions of objects. (3) While natural images are all captured with standard cameras as RGB images, medical images are from \textbf{multi-modal} imaging devices such as \ac{ct}, \ac{mri}, or ultrasound.

\begin{table}[t]
    \centering
    \begin{tabular}{@{}lllll@{}}\toprule
         Dataset&Task& Small & Imbalanced & Multi-modal  \\ \midrule
        RadImageNet & Classification &&\Checkmark&\Checkmark \\ 
        HAM10000 & Classification& \Checkmark&\Checkmark& \\
        MSD Liver &Segmentation& \Checkmark&\Checkmark& \\
         \bottomrule
    \end{tabular}
    \caption{Overview of characteristics of our datasets.}
    \label{tab:dataset_characteristics}
\end{table}
In this study, we aim to give a broad discussion of settings in medical \ac{ai}. Hence, we have chosen three datasets, which encompass the above-discussed scenarios (c.f. Table \ref{tab:dataset_characteristics}).
\begin{enumerate}
    \item 
The RadImageNet dataset \cite{radimagenet} contains over $1.3$ million 2D images with \ac{ct}, \ac{mri}, and ultrasound scans representing three imaging modalities with 165 classification targets, which are highly imbalanced. 
    \item 
    The HAM10000 dataset \cite{tschandl2018ham10000} is a collection of $10\,000$ skin lesion RGB images spread across seven categories. 
We intentionally amplified the class imbalance to a strong but not untypical $80:20$ class ratio by merging classes based on the need for immediate treatment (see Section \ref{sec:appendix_data}).
    \item 
Lastly, we use the MSD Liver dataset \cite{simpson2019large,antonelli2022medical}, a demanding image-to-image task involving just $131$ \ac{ct} scans annotated at voxel level. Given the small number of available training samples as well as a segmentation task (i.e., per-pixel classification) with tumours only encompassing a tiny fraction of each scan, it represents a very challenging medically relevant task.
\end{enumerate}
To the best of our knowledge, no prior work shows the performance of \ac{ai} models trained under formal privacy guarantees on such a comprehensive and large dataset as RadImageNet or a 3D image-to-image task as MSD Liver represents.
\begin{table}[t]
    \centering
    \footnotesize
    \begin{tabular}{lllllll}\toprule
        Privacy Budget & Noise & \multicolumn{2}{c}{Test MCC} & \multicolumn{3}{c}{Reconstruction Risk} \\
        $\varepsilon\textrm{ at }\delta=8.0\cdot10^{-7}$& $\sigma$ &  \multicolumn{2}{c}{Mean $\pm$ StD.} & Worst-Case & Relaxed & Realistic \\\midrule\addlinespace[0.5em]
        \multicolumn{7}{c}{RadImageNet}\\\midrule % $\eta=0.025$
        $1$ &$0.67$& \multicolumn{2}{c}{$64.95\%\pm0.13\%$}& $0.00\%$&$0.00\%$&$0\%$\\
        $8$ &$0.34$&  \multicolumn{2}{c}{$68.75\%\pm0.13\%$}&  $0.04\%$&$0.01\%$&$0\%$\\
        $32$ &$0.267$& \multicolumn{2}{c}{$69.99\%\pm0.25\%$}&  $13.18\%$&$3.96\%$&$0\%$\\
        % $20$ &$\% $&$ \%$ & $\%$ & $0.57\%$&$\%$&$ \%$\\
        % $50$ &$\% $&$ \%$ & $100\%$ & $17.71\%$&$\%$&$ \%$\\
        % $80$ &$\% $&$ \%$ & $100\%$ & $55.05\%$&$\%$&$ \%$\\
        % $100$ & $\% $&$ \%$ &$100\%$&$73.84\%$&$\%$&$ \%$\\
        % $200$ & $\% $&$ \%$ &$100\%$&$97.48\%$&$\%$&$ \%$\\
        $10^{12}$&$0.054$&  \multicolumn{2}{c}{$70.83\%\pm0.19\%$}&$100\%$&$100\%$&$0\%$\\
        Non-private  &$0$&  \multicolumn{2}{c}{$71.83\%\pm1.86\%$}& $100\%$ &$100\%$&$100\%$\\
        \midrule\addlinespace[0.5em]
        \multicolumn{7}{c}{HAM10000}\\\midrule
        % && &&& $\eta=0.005$ \\\midrule
        $1$ &$0.92$& \multicolumn{2}{c}{$15.60\%\pm4.13\%$}& $0.03\%$& $0.01\%$ &$0\%$\\
        $8$ & $0.47$& \multicolumn{2}{c}{$37.48\%\pm3.45\%$}& $1.22\%$& $0.04\%$ &$0\%$\\
        $20$ &$0.40$& \multicolumn{2}{c}{$42.83\%\pm2.37\%$}& $22.30\%$& $0.78\%$ &$0\%$\\
        % $100$ & $45.10\%$&$3.46\%$ &$100\%$&$100\%$&$0\%$\\
        % $500$ & $50.15\%$&$2.95\%$ &$100\%$&$100\%$&$0\%$\\
        % $2\,000$ & $51.30\%$&$3.63\%$ &$100\%$&$100\%$&$0\%$\\
        $10^{9}$&$0.02$&  \multicolumn{2}{c}{$51.98\%\pm2.52\%$}& $100\%$&$100\%$&$0\%$\\
        Non-private  & $0$& \multicolumn{2}{c}{$51.66\%\pm1.38\%$}& $100\%$  &$100\%$&$100\%$\\
        
        \midrule\addlinespace[0.5em]
        \multicolumn{7}{c}{MSD Liver}\\\midrule
        && Dice Score Liver& Dice Score Tumour  & \multicolumn{3}{c}{Reconstruction Risk}\\
        % & \multicolumn{2}{c}{Liver} & \multicolumn{2}{c}{Tumour}& Optimal & GLRT & Optimal & GLRT & Empirical \\
        &&Mean $\pm$ StD.&Mean $\pm$ StD. & Worst-Case & Relaxed & Realistic \\\midrule
        $1$ &$9.97$&$42.84\%\pm1.83\% $&$0.96\%\pm0.37\% $& $1.66\%$ & $0.97\%$ & $0\%$\\
        $8$ &$1.66$&$74.71\%\pm3.14\%$&$3.01\%\pm0.96\% $& $17.96\%$ & $3.68\%$ & $0\%$\\
        $20$ &$0.96$&$79.06\%\pm2.17\% $&$5.55\%\pm0.72\% $& $74.24\%$ & $27.37\%$ & $0\%$\\
        $10^{9}$ &$0.0054$&$91.20\%\pm0.23\%$&$29.73\%\pm2.89\% $& $100\%$ & $100\%$ & $0\%$\\
        % $1 \times 10^{18}$ &$1.71\cdot10^{-7}$&$92.62\%\pm0.17\% $&$28.11\%\pm2.89\% $& $100\%$ &$100\%$& $0\%$\\
        Non-private &$0$&$91.58\%\pm0.41\% $&$28.38\%\pm2.29\% $&$100\%$ &$100\%$& $100\%$\\
        \bottomrule
    \end{tabular}
    \caption{\textbf{Comparison of performance to privacy risk over multiple datasets and privacy budgets.} 
    Test MCC denotes Matthew's Correlation Coefficient on the test dataset. For all performance metrics, we give the mean $\pm$ the standard deviation (StD.) over five runs with different random seeds. Reconstruction risk denotes the upper bounds for the risk of a successful reconstruction attack of a worst-case and minimally relaxed adversary, as well as the empirical success of one of the strongest \protect\say{realistic} attacks. An image is considered successfully reconstructed if the \acf{ssim} to any reconstruction is higher than $80\%$. Note that the noise multiplier $\sigma$ is given for the empirical attack scenario where an adversary manipulated hyperparameters in their favour. Noise multipliers for performance analysis are generally higher.
    }
    \label{tab:results}
\end{table}
\begin{figure}
    \centering    
    \begin{subfigure}{0.3\textwidth}
        \includegraphics[width=\textwidth]{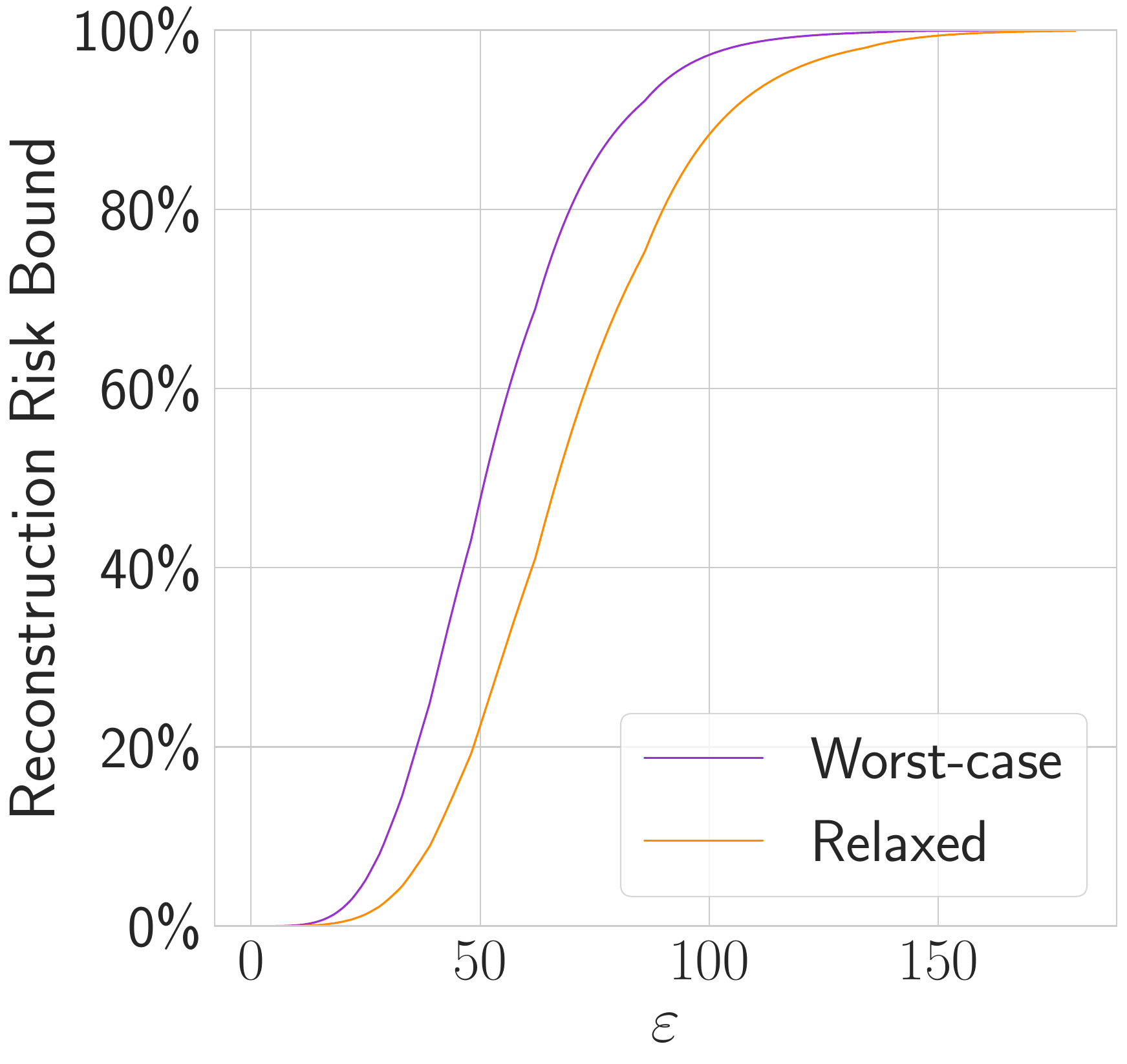}
       % \caption{Image 1}
        \label{fig:radimagenet_rero}
    \end{subfigure}
    \begin{subfigure}{0.3\textwidth}
        \includegraphics[width=\textwidth]{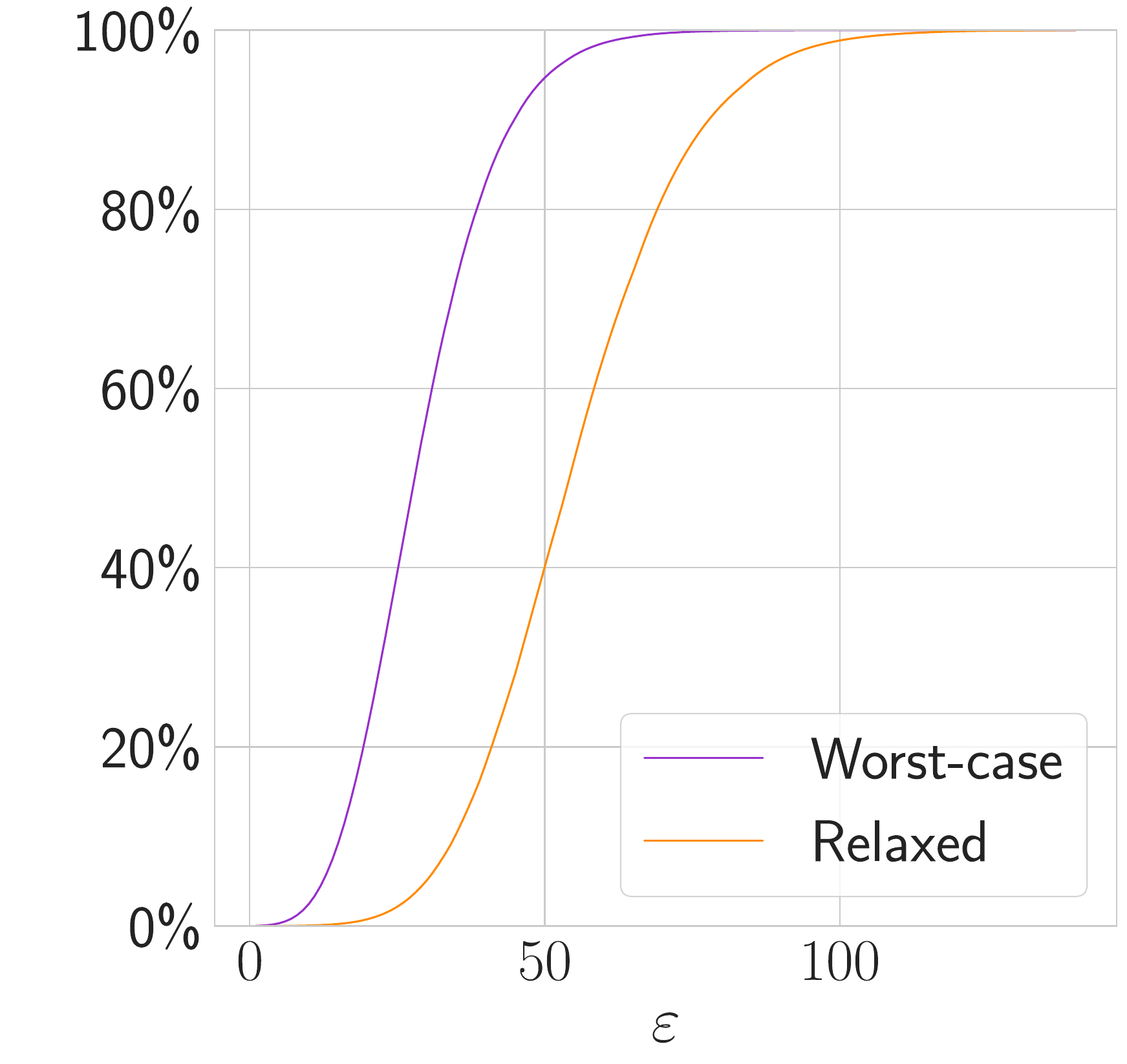}
        % \caption{Image 1}
        \label{fig:ham_rero}
    \end{subfigure}     
        \begin{subfigure}{0.3\textwidth}
        \includegraphics[width=\textwidth]{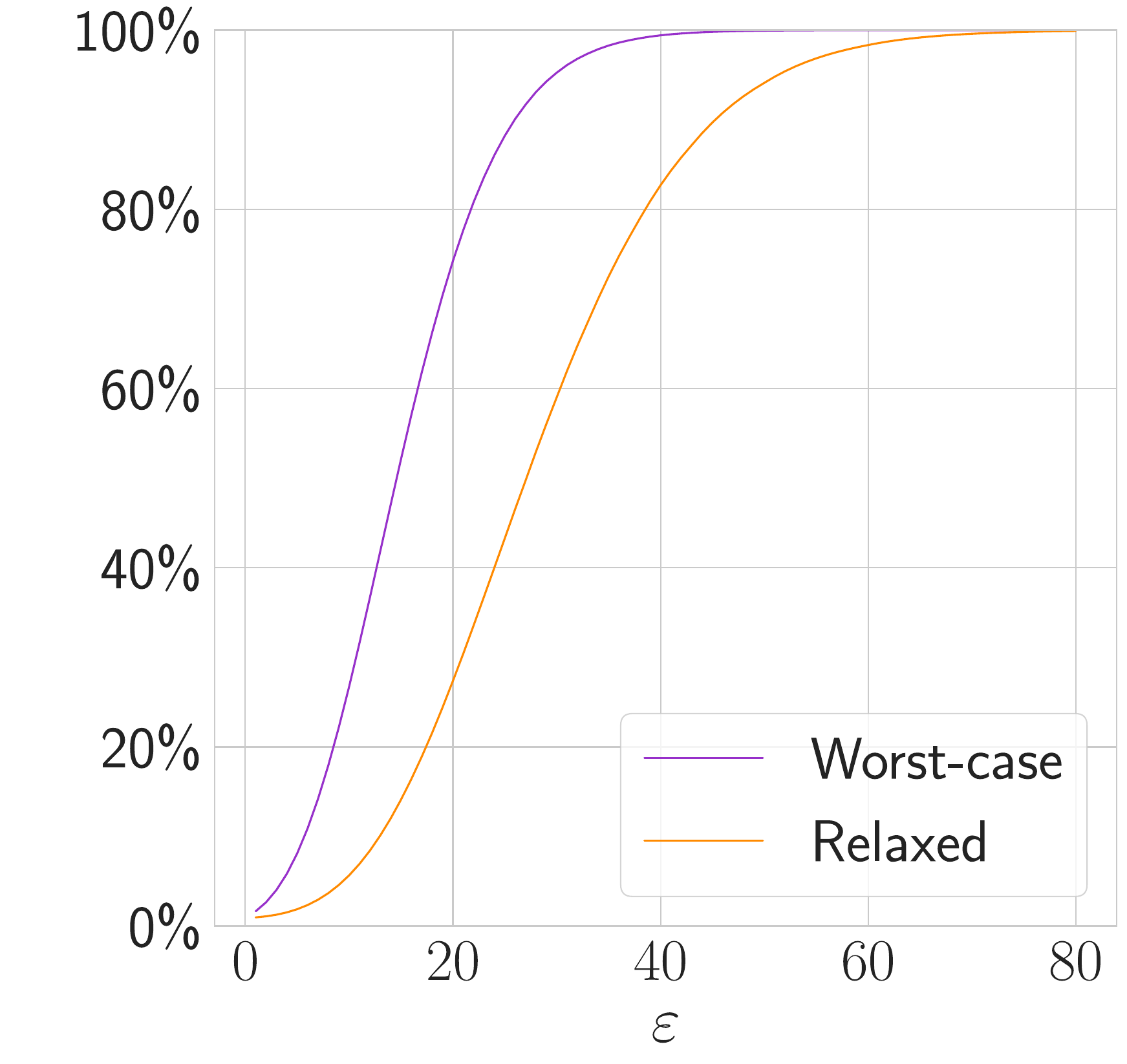}
        % \caption{Image 1}
        \label{fig:msd_rero}
    \end{subfigure}
    \caption{\textbf{Theoretical reconstruction bounds for a worst-case and slightly relaxed adversary.} From left to right: RadImageNet, HAM10000, MSD Liver. We see that the mathematical upper bound for a reconstruction risk of a minimally relaxed threat model (orange) is already substantially lower compared to a worst-case setting (purple).}
    \label{fig:theoretical-bounds}
\end{figure}
\def\figheight{0.25\textheight}
\def\curveheight{3.5cm}
\begin{figure}
    \centering
    % Dataset 1

    % \begin{subfigure}{0.45\textwidth}
    %     \includegraphics[height=\figheight]{figures/radimagenet/characteristic_rero_curve_rad.pdf}
    %     % \caption{Image 2}
    %     \label{fig:radimagenet_cac}
    % \end{subfigure}
    % \begin{subfigure}{0.45\textwidth}
    %     \includegraphics[height=\figheight]{figures/radimagenet/reconstructions_ordered_rad.pdf}
    %     % \caption{Image 3}
    %     \label{fig:radimagenet_recon}
    % \end{subfigure}
    
    % % Dataset 2

    % \begin{subfigure}{0.45\textwidth}
    %     \includegraphics[height=\figheight]{figures/ham/characteristic_rero_curve_ham.pdf}
    %     % \caption{Image 2}
    %     \label{fig:ham_cac}
    % \end{subfigure}
    % \begin{subfigure}{0.45\textwidth}
    %     \includegraphics[height=\figheight]{figures/ham/reconstructions_ordered_ham.pdf}
    %     % \caption{Image 3}
    %     \label{fig:ham_recon}
    % \end{subfigure}
    
    % % Dataset 3

    % \begin{subfigure}{0.45\textwidth}
    %     \includegraphics[height=\figheight]{figures/msd/characteristic_rero_curve_msd.pdf}
    %     % \caption{Image 2}
    %     \label{fig:msd_cac}
    % \end{subfigure}
    % \begin{subfigure}{0.45\textwidth}
    %     \includegraphics[height=\figheight]{figures/msd/reconstructions_ordered_msd.pdf}
    %     % \caption{Image 3}
    %     \label{fig:msd_recon}
    % \end{subfigure}
    \includegraphics[width=0.95\textwidth]{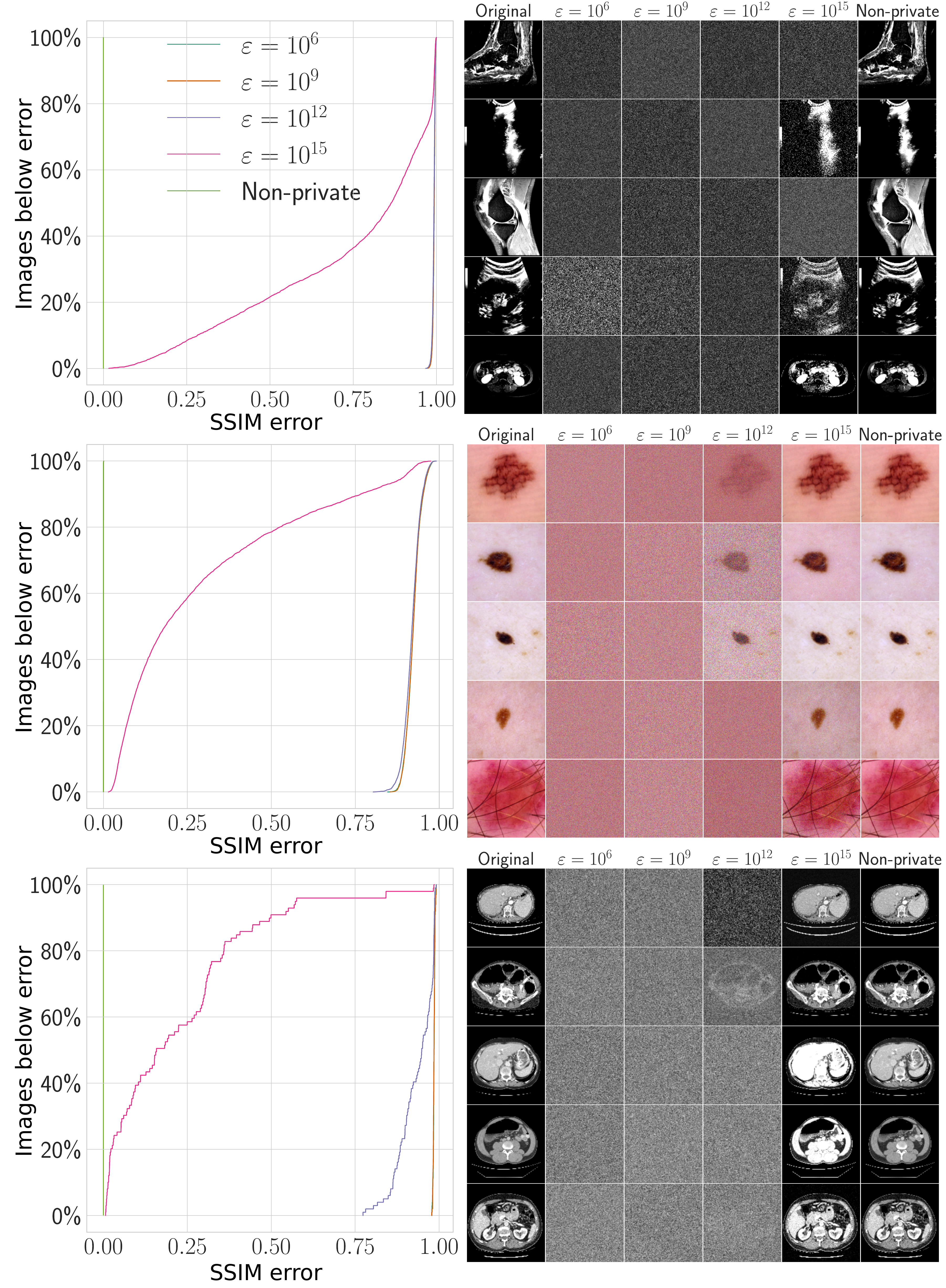}
    
    \caption{\textbf{Reconstruction threat analysis for three datasets.} Each row shows one dataset. From top to bottom: RadImageNet, HAM10000, MSD Liver. Left column: Cumulative number of images, which have in an empirical reconstruction a structural similarity (SSIM) difference lower than the value on the x-axis. Note that it is the SSIM reconstruction error and thus perfect at 0 and worst at 1. Exemplarily, we see that on the MSD dataset at a reconstruction error of $10\%$ all non-private (green) images, $39\%$ at $\varepsilon=10^{15}$ (pink) and none at more restrictive privacy guarantees can be reconstructed
    On the right: The top five images with the best reconstruction score and their corresponding best reconstruction at various privacy budgets.}
    \label{fig:reconstruction}
\end{figure}

\subsubsection{Evaluation}
Our evaluation focuses on how various privacy risks on multiple datasets correlate to the algorithm's performance. First, we show the correlation of the \ac{ai} performance on our datasets to privacy budgets. Second, we illustrate the implications of a certain privacy budget in a risk profile, summarising the reconstruction risk under different threat models. We recall that a threat model corresponds to the set of assumptions over the attacker, where we give the theoretical bounds for a worst-case and a slightly relaxed adversary. Both are more pessimistic than any real-world scenario. Thus, we add a third threat model representing the worst \say{realistic} case.  

To measure the performance of the models on classification tasks, we use \ac{mcc} \cite{matthews1975comparison}. 
Opposed to more frequently used metrics such as accuracy or $F_1$-score, it incorporates the entire confusion matrix and, by that, is extremely robust against any class imbalance \cite{chicco2020advantages}. 
It is also better interpretable as for random predictions it is $0$ and for perfect predictions $1$, whereas the accuracy depends on the class distribution. 
For the segmentation task, we measure the class-wise Dice score of the 3D volumes and report the average over all volumes for the liver and tumour, as they are the targets of interest in our task. A perfect prediction yields a $100\%$ Dice score.

In Table \ref{tab:results} we list for all datasets over increasing privacy budgets the best possible \ac{ai} model performance and the corresponding reconstruction risk. The risk is three-tiered: (1) The upper bound of a \textbf{worst-case} adversary. This is the absolute maximum that the risk can be under this setting and can not be increased by post-processing or side information. (2) The upper bound of a minimally \textbf{relaxed} adversary as introduced by \cite{kaissis2023bounding}. (3) The reconstruction success of the \textbf{real-world} adversary. We argue that for practical use cases, protection against such an attacker suffices. By listing all three, we provide an overview of how the risk varies in constant settings by changing assumptions about the adversary.
% We contrast the \ac{ai} performance with the theoretical upper bound of successful data reconstructions of a worst-case and minimally relaxed adversary \cite{kaissis2023bounding}, as well as the empirical reconstruction success of a \ac{sota} reconstruction attack in the most vulnerable \say{realistic} setting. An overview can be seen in Table \ref{tab:results}.

\subsection{Performance Trade-Offs Under Varying Privacy Levels}
\subsubsection{Impacts on Model Performance is Substantial for Small Datasets}
At first, we analyse the impact of a very restrictive (small) privacy budget of $\varepsilon=1$ on the \ac{ai} predictive performance of our datasets (see Table \ref{tab:results}). 
Across the board, we see that at these budgets, the impacts on the model performance are strong. 
Concretely, we find that on RadImageNet, a standard non-private \ac{ai} model reaches $71.83\%$ on average, while trained at such restrictive privacy guarantee we find an average \ac{mcc} of $64.95\%$, which is still $90\%$ of the non-private \ac{mcc} score. 
The gap turns much larger on the HAM10000 dataset, where the model performance, when trained with a very low privacy budget of $\varepsilon=1$ is closely above the chance level at an \ac{mcc} of $15.60\%$. 
Similarly, on the MSD Liver dataset at restrictive privacy budgets, the average Dice score for the liver goes down to $42.84\%$ (Non-private: $91.58\%$) and negligible $0.96\%$ for the tumour.
This exemplifies the challenges of furnishing strong privacy protection when training \ac{ai} models on small or difficult datasets.

\subsubsection{Prediction Quality under Medium Budgets Depends on Dataset}
Next, we consider medium privacy budgets that range in our study from $\varepsilon=8$ to $\varepsilon=32$, which are a typical choice \cite{de2022unlocking,sander2022tan} in literature. 
As $\varepsilon$ is an exponential parameter ($e^\varepsilon$), larger values correspond to exponentially increased privacy guarantees. 
For this reason, some argue that the guarantees provided by such medium budgets are meaningless  \cite{dwork2014algorithmic,stock2022defending}.

At these privacy budgets, although the performance substantially increases, compared to the extremely restrictive privacy budget, on none of the datasets, the private \ac{ai} models match the non-private performance. 
On the RadImageNet dataset, the achieved result closely approaches the non-private baseline. 
Specifically, with a privacy budget of $\varepsilon=32$, the \ac{mcc} score is $69.99\%$ compared to $71.83\%$ of the corresponding non-private performance.
Also for the HAM10000 dataset performance is now strongly improved at $42.83\%$ \ac{mcc} score, yet is still a decrease of $9\%$ compared to the non-private result.
Lastly, in the MSD Liver dataset, the liver as a larger organ can now be learned up to a reasonable dice score of $79.06\%$ at $\varepsilon=20$. 
However, it is also still far from the non-private performance. 
Especially concerning is the prediction quality of the tumour, which is a much smaller and more complex structure. 
This leads to a poor segmentation quality and only achieves, on average, a dice score of $5.55\%$, which is unsuitable for a real-world application.
Again, we note that performance trade-offs impact especially smaller and imbalanced datasets.

\subsubsection{Performance Trade-Offs Vanish Under Large Privacy Budgets}
For very large privacy budgets, we observe that the gap between private and non-private performance disappears. 
We recall that HAM10000 and MSD Liver as small datasets are extremely challenging under restrictive \ac{dp} conditions. When increasing the privacy budget to $\varepsilon=10^{9}$ no statistically significant difference to the non-private model can be detected anymore (p-values: HAM10000: $0.36$, MSD Liver dataset p-value liver: $0.10$, tumour: $0.29$, Student's t-test).
%On MSD liver at a privacy budget of $10^{18}$, the private model is even significantly better (p-value: 0.002) in predicting the liver. We suspect this to be caused by a small regularising effect, which gradient clipping imposes \cite{zhang2020gradient}. 
Only on RadImageNet, although the non-private model is still statistically significantly superior (p-value: 0.001), the private model at an $\varepsilon=10^{12}$ achieves $99\%$ of the non-private baseline performance.
%However, we note that we limited our experiments to two model architectures that are known to perform well with differentially private training and cannot rule out that for example large vision transformers could further boost the non-private performance. 

It is unsurprising that increasing the privacy budget lifts the implications on the model performance. 
Hence, the question that must be asked is what level of privacy is necessary for a specific setting. 
This cannot be answered generally and must be carefully considered for each use case. 
Important for these considerations is which risks are associated with a certain privacy budget. 
We analyse the reconstruction risk for our three datasets and three threat models discussed previously. 

\subsection{Low Worst-Case Reconstruction Bounds Require Small Privacy Budgets}
Although, for most use cases too pessimistic, worst-case analyses have the advantage that we have a formal guarantee, which gives an absolute upper bound on the risk of this scenario. 
When analysing the theoretical worst-case (highest) success of reconstruction attackers, we find that for the large RadImageNet dataset for budgets $\varepsilon\leq8$, the risk is smaller than $0.05\%$. 
However, already at an $\varepsilon=32$ almost $15\%$ of the samples can theoretically be recovered. 
Here, the smaller datasets are again at higher risk. 
While at $\varepsilon=1$ the risks remain low, it strongly increases at $\varepsilon=8$  for HAM10000 ($0.03\%$ to $1.22\%$) and MSD Liver ($1.66\%$ to $17.96\%$).
At $\varepsilon=20$ theoretically, up to $74.24\%$ of all data samples of the MSD Liver dataset can be reconstructed. 

However, even minimally relaxing the assumptions for such an adversary to a threat model decreases the risk associated with these privacy budgets drastically. 
We recall that under this threat model, the only thing which is relaxed compared to the worst-case is that the attacker does not know the sample, which is reconstructed beforehand. 
Yet, for theoretical analysis, there is still the assumption that the reconstruction algorithm is either perfect or fails and the risk which is then calculated is the maximum rate where the attacker correctly decides if the reconstruction they obtained was indeed the dataset sample in question.
This threat model is still too pessimistic for any real-world use case and the analysis is mostly for theoretical purposes. 
Still, such a practically irrelevant minimal relaxation already gives a much more favourable risk profile, especially for medium privacy budgets. 
Exemplarily, the risk associated with $\varepsilon=20$ goes down from over $20\%$ to less than $1\%$ for the HAM10000 dataset. 
Similarly, the risk for the MSD dataset at $\varepsilon=8$ decreases from $18\%$ to $4\%$. 
A visualisation of the risk difference in worst-case and relaxed threat models can be found in Figure \ref{fig:theoretical-bounds}.

\subsection{Empirical Protection is Provided Even at Large Privacy Budgets}
The previously discussed theoretical analyses show rapidly growing risks associated with small and medium privacy budgets. 
However, as discussed before, we argue that these analyses are too strict for any \say{realistic} use-case.
Hence, we ask what is the worst case of any practical scenario. 
We find this as a federated learning setup, where a central server coordinates the learning on the data of distributed clients, which follow each training command sent by the server. 
This implies that the server can freely choose any network architecture and hyperparameters. 
Note that any client who performs a simple check would notice such a malicious server. 
For such cases, attacks have been shown in literature, which analytically can recover the model input perfectly \cite{fowl2021robbing,boenisch2023curious}. 
We employ these attacks as empirical risk assessments. 
% Yet, as they are based on strong assumptions, we added an empirical analysis of the worst \say{realistic} scenario using state-of-the-art attacks. 
For our analysis to measure the reconstruction success, we use the structural similarity score, which is a standard metric for image similarity \cite{wang2004image}. 

In contrast to the aforementioned theoretical risk bounds, we find that for practical attacks even so far considered meaningless privacy budgets $\varepsilon>10^9$ can provide effective protection against successful reconstruction. 
In the left column of Figure \ref{fig:reconstruction}, we visually demonstrate for several privacy budgets the step curve of how many dataset images are below an increasing \acf{ssim} error.
It can be thought of as the cumulative density function of the distribution of reconstruction errors.
We observe that for all datasets without the addition of \ac{dp} constraints, nearly all images can be reconstructed perfectly. 
As soon as \textit{some} privacy guarantee is introduced, even very generous budgets at an $\varepsilon$ of one billion provide empirical protection against the reconstruction of data samples. 
Furthermore, confirming previous works \cite{fowl2021robbing,usynin2022zen}, our threat model is still extremely powerful. 
A server without the control of hyperparameters but still over the model architecture already imposes a substantially lower reconstruction risk. 
If the server does not set the batch size to one but is set to the real training batch size, for example, on the RadImagenet dataset even in the non-private case we could only reconstruct less than $5\%$ of all images at a batch size of $3328$.  %then for MSD Liver we could still reconstruct all images perfectly (batch size: 2), while for
We note that such large privacy budgets we found to still protect are so far universally shunned by the privacy community as meaningless. 
In fact, it is remarkable as $\varepsilon$ is an exponential parameter, implying that $\varepsilon=10^{9}$ is $e^{10^{9}-1}$ times \say{less private} than $\varepsilon=1$. 
This number is almost inconceivably large and, for example, \textit{much} larger than the number of atoms in the universe.
Yet, we show that even this pinch of privacy has drastic effects if practical scenarios instead of worst-case analyses are considered. 
Complemented by the above-discussed finding that performance trade-offs melt away in these minimally restrictive realms, this signifies a potential compromise between protection and real-world usability.

%In contrast to the theoretical risk bounds, which grow rapidly with higher privacy budgets, we found that empirically for the protection against a \ac{sota} reconstruction attack even humongous privacy budgets still provide effective protection. This is especially striking in the case of the MSD Liver dataset, where in the standard non-private setting, we could reconstruct every single training image with \acf{ssim} \cite{wang2004image} of less than $\eta=0.25$. However, at an previously considered meaningless privacy budget of $\varepsilon=10^{18}$ no training sample was below this reconstruction error threshold. This is further visualised in Figure \ref{fig:reconstruction}.

\section{Discussion}
In this study, we explore the relationship between privacy risks and \ac{ai} performance in the context of sensitive applications, such as medical imaging. 
Currently, practitioners are confronted with a trade-off triangle where \ac{ai} performance, privacy protection, and computational efficiency are conflicting, and no solution has so far been found to facilitate all of these goals.
Previous work showed that pretraining on a massive dataset of four billion images could allow models to transfer to private datasets \cite{berrada2023unlocking}.
However, in practice this is typically not feasible due to limited access to such large datasets or the computational resources to train such a model.
Furthermore, such data scales only exist for natural 2D images but not yet for 3D images, which are typical in medical imaging. 
Therefore, often the choice remains for practitioners to prioritise privacy and sacrifice performance or to put entrusted sensitive data at risk of being leaked. 
Currently, there is no clear method to balance these two objectives, leaving practitioners without guidance. To make informed decisions on how to trade off these competing interests, a broad discourse involving ethicists, lawmakers, and the general population is crucial. 
A prerequisite of this dialogue is understanding the risks associated with specific privacy budgets and the potential trade-offs in \ac{ai} performance. 
Our comprehensive study across three representative medical imaging datasets with different characteristics lays the foundation for this conversation.
We find that a significant category of risks can factually be averted without this trade-off. 
In fact, privacy-performance trade-offs have so far always been based on worst-case assumptions, which do not overlap with realistic training settings.
In most real-world scenarios, an attacker will not already have access to the data, as they could simply use the dataset to obtain the information in that case. 
Furthermore, because of its maximal simplicity usually considered inference membership attack of an already available data point on an \ac{ai} model gives the attacker only minimal additional information.
We postulate that it is more critical to prevent the reconstruction of sensitive data in the worst real-world setting. 
We show that for derisking the threat of reconstruction attacks in realistic settings large privacy budgets suffice. 
Even more, we find that the trade-off between privacy risks and model performance vanishes when using such large but protective privacy budgets.
This compromises privacy risks, model performance and computational efficiency and could allow a widespread breakthrough for the use of \ac{dp} in medical settings.

From previous studies \cite{kaissis2021end,ziegler2022defending,balle2022reconstructing,hayes2023bounding}, we already know that privacy-enhancing techniques protect \ac{ai} models in sensitive contexts from reconstruction attacks. 
We found that even considerably relaxed privacy budgets can offer substantial protection against data reconstruction attacks in realistic settings. 
In such scenarios, the \ac{ai} performance loss is negligible. 
While we note that our results are based on an empirical analysis, it is apparent that \ac{dp} training with minimal guarantees still provides better protection than non-private \ac{ai} training. 
Considering this finding that we can have \textit{some} protection without strong decreases in model performance, it seems negligent to train \ac{ai} models without any form of formal privacy guarantee. 
% We point out that there is a potential gap between theoretical analysis and real attacks. 
We note, that for all intents an purposes our empirical attack can be considered sufficient for all current attack scenarios. 
However, as we have no theoretical lower bound, the possibility exists that future attacks could achieve success closer to the upper bound.
Because of this, we explicitly warn readers to take our results as a \textit{carte blanche} to use arbitrarily high privacy budgets. 
The truth lies in the middle: If the alternative is to not use any privacy at all, rather use \ac{dp} with a very high budget. 

On this note, we remark that the effectiveness of the \ac{dp} protection against attacks is a factor of the noise multiplier at a constant clip norm, batch size, duration of the training and the training set size. 
% This is important, as the noise multiplier does not exclusively depend on the privacy budget, meaning, two \ac{dp} algorithms with the same privacy budget could exhibit a different susceptibility to reconstruction.
% For example, a training with just a single training step would have a much lower noise multiplier than at the same budget compared to a real training with typically several thousand steps. 
However, as shown, these noise multipliers are small enough to be converted to large privacy budgets and facilitate high-performing \ac{ai} models. 
We also observed that the \ac{ai} performance loss introduced by \ac{dp} tends to be smaller on larger datasets, due to less injected noise per sample and more information to achieve a certain privacy budget at consistent hyperparameters. 
Yet, many medical datasets are inherently small and as seen above strong privacy protection can prevent successful learning of high-quality tumour segmentations. This can have negative consequences for the applicability of such networks in clinical practice. 
For models to be effectively trained on such challenging datasets, when pretraining is not possible for reasons of data availability or computational resources our techniques come to a limit and there might be a need to either accept elevated privacy risks or obtain access to more data.
The solution to both problems might go hand in hand as with more robust and unbreakable mathematical guarantees safeguarding data privacy, we anticipate that patients may be more inclined to share their data, thereby allowing large-scale medical AI training.
If such a scenario occurs the privacy-performance trade-offs presented might even be more favourable than our findings indicate. 
This would be complemented by a workflow, where multiple \aclp{pet} are employed to enable various aspects to privacy. 
For example, a system using Federated Learning to assert the data governance remains at the original hospital, Secure Aggregation to conceal contributions from different sites, and \acl{dp} to limit the private information of single patients demonstrated in previous works \cite{kaissis2021end} would provide a holistic workflow.

We note that our choice of datasets and architectures is motivated by medical imaging settings.
In those settings, typically computational resources are limited and data is scarce.
In fact, we are convinced that the widespread use of such methods will only ensue once they can be used by the majority of practitioners who typically are without access to large computing clusters. 
Hence, we carefully designed our study to cover typical and representative medical problems to provide a holistic analysis with trade-offs in computational resources.
Under these considerations, we limited ourselves to a few model architectures, which are known to be trained efficiently (ResNet, DenseNet, UNet) and datasets, which represent a broad range of typical problems. 

An additional technical limitation stems from the fact that the authors of the RadImagenet dataset \cite{radimagenet} mention, that some patients contributed multiple images. 
However, we have no information about image-to-patient correspondence. 
As we calculate the privacy guarantees over the dataset per image, the per-patient privacy guarantee depends on the number of images one patient contributed and might be significantly lower. 

%In conclusion, we hope to provide a basis for a wide discussion on how to decide on a specific privacy budget for various scenarios. We are not aware of any current legislation, which provides practitioners with concrete guidelines, beyond the call for risk moderation. We show that in contexts, where sensitive data is handled not using any form of privacy preservation is negligent.
In conclusion, we show that even the use of very loose privacy guarantees still provides substantially better protection than standard \ac{ai} training, while achieving comparable performance. This can be the compromise between provable risk management and performance trade-offs, which previously prevented the breakthrough of \ac{dp}. Further research should be directed towards analysing various threat models such that not only worst-case scenarios are considered. Only by illuminating the risks for multiple scenarios the basis for a wide discussion among ethicists, policymakers, patients and other stakeholders is provided, on how to trade-off privacy and performance as fundamental goals of \ac{ai} in sensitive applications.

%TC:ignore
\newcommand{\Pdiv}{\mathbin{\text{$\vcenter{\hbox{\textcircled{$/$}}}$}}}

%TC:ignore
\section{Methods}
In this section we report all details necessary for our experiments on training models in a differentially private way on our datasets as well as the procedures to analyse risk profiles. Furthermore, we describe the rationale for several choices in our study design and describe hyperparameters necessary for reproducibility. 

\subsection{Data}\label{sec:appendix_data}
In Section \ref{sec:datasets} we described characteristics of typical medical datasets. 
We note, that these characteristics partially amplify the negative performance impact by the constraints introduced by \ac{dp}. 
Broadly speaking, at a constant clipping norm the amount of introduced noise during the \ac{dp}-process determines the negative impact on the \ac{ai} performance. 
At any privacy budget the injected noise increases if more training steps are performed or if a higher sampling rate, i.e., the ratio between batch size and dataset size, is used. 
However, the batch size is typically irrespective of the dataset size, which implies that smaller datasets typically have higher sampling rates. 
Furthermore, they often require more training epochs, i.e., the amount of times the entire dataset was (on average) presented to the network. 
As a consequence, the amount of noise that is injected when training on small datasets compared to larger ones is increased and higher performance penalties are expected.
Furthermore, \ac{dp} bounds the magnitude a single sample can change the training. 
This is important for training with imbalanced datasets with underrepresented classes, which often suffer an additional performance loss \cite{bagdasaryan2019differential}.

For detailed descriptions of the datasets we refer to the original publications \cite{radimagenet,tschandl2018ham10000,simpson2019large,antonelli2022medical}. In the following, we describe modifications we performed and the effects on the data distribution. 

For the HAM10000 dataset \cite{tschandl2018ham10000} we merged classes into whether there is indication for immediate treatment, which is still a medically important distinction. By this we convert the multi-class classification problem into a highly imbalanced binary classification problem. We categorised here as follows: 

\begin{tabular}{ll}
    \toprule
    \multicolumn{2}{c}{Treatment indication} \\
     immediate & not immediate \\\midrule
    actinic keratoses and intraepithelial carcinomas & melanocytic nevi \\ 
    basal cell carcinomas & benign keratinocytic lesions \\
    melanomas & dermatofibromas \\
     & vascular lesions \\\bottomrule
\end{tabular}

In total, this dataset has $10\,015$ images, of which $1\,954$ are labelled for immediate treatment and $8\,061$ are not. 
\subsection{Model Training}\label{sec:train}
All of our experiments were performed using an NAdam optimiser, which is extremely robust to learning rate changes allowing us to keep a consistent learning rate of $2\mathrm{e}-3$. Input data was always normalised with the mean and standard deviation of all images in the training set. 
For each dataset, we perform a hyperparameter search, where we evaluate for one privacy level ($\varepsilon=8$) and the non-private training the optimal setting for architecture, batch size, loss weighting, and augmentation. In the non-private case, we perform an early stopping strategy to determine the number of epochs. In the private case, this is not possible as the number of epochs directly influences the amount of added noise. However, previous works showed that longer training almost always yields better results \cite{de2022unlocking}. Yet to limit training time we also search for the point of saturation. Also for reasons of computational complexity, we assume that the optimal settings for these parameters transfer to all other privacy regimes. Furthermore, we limit the choice of architectures to a ResNet-9 with ScaleNorm and a WideResNet40-4, which have in previous literature been proven to be especially suited for differentially private training \cite{klause2022differentially, de2022unlocking}. In the segmentation case, we limit ourselves to a standard U-Net \cite{ronneberger2015u, cciccek20163d}, where we optimise the number of channels on the bottleneck. We then evaluate for each privacy setting separately the optimal clipping norm. Again for reasons of computational complexity, we evaluate this after one epoch and assume it transfers to longer trainings. Finally, we train for each setting five models with different random seeds and report the mean and standard deviation of the respective performance metric. %We use this metric because of it's clear advantages over standard metrics such as accuracy or AUROC \cite{chicco2020advantages}.

All our models are trained from \say{scratch}, i.e., we have not pretrained on any other dataset. This is because there is no \say{good choice} of a dataset for pretraining. ImageNet, which for most computer vision tasks is the standard, is not very effective for medical imaging tasks \cite{radimagenet}. Furthermore, pretraining on medical databases is unacceptable, as it risks leaking the information from the pretraining data, which could be just as private \cite{abascal2023tmi,tramer2022considerations}.

%\subsection{Privacy parameters}
We used the Opacus \cite{Opacus} library for accounting the privacy loss. In particular, we used an RDP accountant, as it provides numerically the most stable implementation. We used an extension of the objax library \cite{objax2020github} as implementation for the DP-SGD algorithm. 

We will open-source the program code used for this study upon acceptance of the manuscript.
\subsubsection{RadImagenet}
As described in section \ref{sec:train}, we analysed the architecture, number of epochs, batch size, loss and multiplicity for the non-private and one private setting ($\varepsilon=8$). For the non-private case, we found a WideResNet40-4 using an unweighted loss function, a batch size of 16, and random vertical ($p=0.2$) and horizontal flips ($p=0.1$) as augmentation to yield the best results. To determine the number of epochs, we used an early stopping strategy with a patience of 5 epochs and $0.1\%$ improvement threshold. For the private case, a ResNet-9 trained for 50 epochs, using an unweighted loss function, using an augmentation multiplicity of 4 again with random vertical ($p=0.2$) and horizontal ($p=0.2$) flips with a batch size of 3328 yielded best results. The clipping norm was tuned for each budget separately and was set as follows: 

\begin{tabular}{lllll}\toprule
    $\varepsilon$ & $1$ & $8$ & $32$ & $1\mathrm{e}12$ \\\midrule
    Clip Norm & $6.46$ & $5.66$ & $5$ & $3.75$ \\ \bottomrule
\end{tabular}

\subsubsection{HAM10000}
For the modified HAM10000 dataset we found the ResNet9 to perform best in private and non-private settings. In the non-private case, we trained with a weighted loss function at a batch size of $32$ using random vertical flips ($p=0.5$) as augmentation. We trained using an early stopping strategy using a patience of $50$ epochs at a minimal improvement threshold of $0.1\%$. For the private case, we used an unweighted loss function at a batch size of 2048 and trained for 100 epochs. We used the same augmentations as in the non-private case for a privacy level of $\varepsilon=10^9$, for all others, we did not use augmentations. Clipping norms are as follows:

\begin{tabular}{lllll}\toprule
    $\varepsilon$ & $1$ & $8$ & $20$  & $1\mathrm{e}9$ \\\midrule
    Clip Norm & $18$ & $8.5$ & $9.5$ &  $9$ \\ \bottomrule
\end{tabular}

\subsubsection{MSD Liver}
For the MSD Liver dataset, we found for both private and non-private cases a U-Net with $16$ channels and no augmentations to perform best. In the non-private case we used a weighted loss function (Background: $0.1$, Liver: $0.4$, Tumour: $0.5$) and trained at a batch size of two. Again, we employed an early stopping strategy with a patience of 50 epochs and a minimal improvement threshold of $0.1\%$. In the private case, we trained at a batch size of one for $500$ epochs. For privacy budgets $\varepsilon\leq 20$ we used an unweighted loss function, for higher privacy budgets we used the same weighting as in the non-private case.

\begin{tabular}{lllll}\toprule
    $\varepsilon$ & $1$ & $8$ & $20$  & $1\mathrm{e}9$ \\\midrule
    Clip Norm & $0.0004$& $0.046$ & $0.0015$ &  $0.33$ \\ \bottomrule
\end{tabular}

\subsection{Reconstruction Risk Analysis}

% \subsection{Evaluating empirical reconstruction success}
In our empirical reconstruction attacks there is no clear way to evaluate whether a specific sample was reconstructed. For each input batch consisting of $N$ samples, we receive $M$ reconstructions. We evaluate this by calculating the pairwise distance between all data samples and reconstructions and assigning each input the reconstruction with the lowest distance. However, this approach loses meaning in the case of images, which have no structure but are entirely dark. This is the case for the RadImagenet dataset, where we put a constraint, that only data samples are considered, which contain more than 10\% non-zero pixels. % Our susceptibility analysis is based on two factors: (1) Theoretical bounds that differential privacy yields, which might be unnecessarily pessimistic for various real-world scenarios, but give a theoretical worst case estimate. (2) The empirical evaluation with a state-of-the-art malicious reconstruction attack to illustrate the protection DP gives at various privacy levels. 

% We compute the theoretical upper bound to the success of membership inference attacks (MIA) and reconstruction attacks \cite{kaissis2023bounding} for two settings. (1) Against an optimal adversary, which has full control and knowledge over the entire training process, model and dataset except the randomness introduced into the mechanism. (2) A minimally relaxed adversary as proposed by Kaissis et al. \cite{kaissis2023optimal} without access to the training database. For the analysis of MIAs the risk can directly be transferred into a receiver-operator curve, which shows the true positive rate (TPR) against the false positive rate (FPR) of an adversary \cite{wasserman2010statistical,dong2022gaussian}. In the case of reconstruction attacks an upper bound on the number of samples where the output of an algorithm can be matched to a specific input in the training data \cite{kaissis2023bounding}. 

We evaluate the practical reconstruction success by using a principle demonstrated in previous literature \cite{fowl2021robbing,boenisch2023curious} adapted to our use case. The network architecture is slightly modified by prepending two linear layers in front of the actual network architecture. The first takes all input image pixels as input and projects them to an intermediate representation of $N$ bins. In our experiments, we set $N=10$. This intermediate representation is afterwards projected again to the number of all pixels and resized to the original image shape. To each of the outputs, the mean of the intermediate representations is added. Afterwards, it can be processed as usual by the remaining neural network. Because our adversary is assumed to have control over all hyperparameters they can set the batch size to one and by that enforce that no reconstruction of two images overlap. If now a gradient is calculated over the network, which is non-zero for the weights $W_i$ and biases $b$ of the first linear layer the input $x$ can be analytically recovered by $x = \nabla_{W_i}\mathcal{L} \oslash \frac{\partial \mathcal{L}}{\partial b}$, where $\oslash$ is the elementwise division. We note that for this attack it is irrelevant what network architecture comes after this imprint block. We used implementations provided by \cite{geiping_github}.
% We note that such a malicious adversary is a strong assumption and might in practice be unnecessarily pessimistic.  We report based on the reconstruction robustness formulation of \cite{balle2022reconstructing} the ratio of reconstructions $\eta$ of which the perceptual distance to a ground truth image is smaller than $\gamma$. We visualize $\eta(\gamma)$ for varying privacy levels, which gives a threshold independent identification of the risk profile.

\subsection{Choice of privacy budgets.}
% As outlined before, a privacy budget $\varepsilon$ can directly be transferred to the optimal ROC-curve an adversary can achieve in a membership inference game under the full DP threat model (see Figure \ref{fig:DP_threat_analysis}). Especially interesting is the true positive rate (TPR) at a minimal false positive rate (FPR), as an attacker at these thresholds can assume that positively classified samples were indeed used for the training of an algorithm \cite{carlini2022membership}. Hence, we report the TPR at an FPR, which was arbitrarily set to $0.2\%$. As these values are independent of the dataset and hyperparameters, we choose for all datasets to incorporate the privacy budgets of $\varepsilon=1$ and $\varepsilon=8$, which are also often used in literature \cite{de2022unlocking, klause2022differentially, tramer2020differentially}. 

For our experiments on the utility trade-off we chose several privacy budgets. We note that this choice was arbitrary. For all experiments, we used a $\delta=8\cdot 10^{-7}$. For all settings, we evaluated $\varepsilon=1$ and $\varepsilon=8$, which are standard values in literature \cite{klause2022differentially,de2022unlocking,sander2022tan}.
Furthermore, we calculate the theoretical reconstruction bound of the worst case and relaxed threat models. As the already included privacy budgets at $\varepsilon=1$ and $\varepsilon=8$ already showcase very low reconstruction bounds, we add one more privacy level for all datasets, where a significant amount of samples is already at risk of being reconstructed. 
In addition, we report a privacy budget $\varepsilon=10^{3N}, N\in\mathbb{N}$, where the characteristic reconstruction robustness curve is still similar to random noise.

%TC:endignore

% Furthermore, we see that for very large datasets the privacy-utility trade-offs are not as drastic as for smaller or imbalanced datasets. For the RadImagenet dataset we achieve results close to the non-private case already at an $\varepsilon=8$. For these datasets the challenge rather lies in the computational overhead of DP-SGD training, where per-sample gradients need to be calculated.
% \section*{Contributions}
% % AZ conceptualised this study, wrote the program code, performed all experiments and prepared the manuscript. TTM assisted in the preparation of the manuscript. SS assisted in the design of the program code. LF wrote program code for an efficient reconstruction matching and segmentation loss. JB helped to prepare the HAM10000 dataset for our purposes. RB and DR provided oversight. GK helped conceptualise this study and in the preparation of the manuscript, wrote code for the theoretical risk bounds and provided oversight. All authors revised the manuscript.
% [Redacted for double-blind review]
\section*{Competing interest}
The authors declare no competing interest. 
\section*{Data Availability Statement}
All datasets used in this study are published and publicly available. Access to RadImageNet \cite{radimagenet} must be requested at \url{https://www.radimagenet.com/}. The HAM10000 dataset \cite{tschandl2018ham10000} is available at \url{https://doi.org/10.7910/DVN/DBW86T}. The MSD Liver dataset \cite{simpson2019large,antonelli2022medical} is available at \url{http://medicaldecathlon.com/}.
\section*{Code Availability Statement}
Our program code will be publicly released upon acceptance of this manuscript. Prior use can be requested via inquiry to the authors. %A blinded version is attached for the purpose of the reviews.
\section*{Environmental impact}
Lastly, we would like to give a rough estimate of the climate impact of this study. We assume the average German power mix which as of 2021 according to the German Federal Environment Agency corresponds to 420g CO$_2$/kWh. Only the final RadImagenet trainings (no hyperparameter optimisation) ran on 8 NVIDIA A40, where we assume a power consumption of 250W on average, each for almost 4 days, 5 privacy levels, and 5 repetitions. Hence, this amounts to around 960kWh and thus more than 400kg of CO$_2$ equivalents. This almost equals a return flight from Munich to London. Hence, we tried to limit our hyperparameter searches to the necessary. %as much as possible to only consider two architectures for each setting, limiting the augmentation multiplicity to four and extrapolating the results after one epoch to longer trainings (which might not be true). 
In total, we assume that this study produced at least 2 tons of CO$_2$ equivalents.

\bibliographystyle{unsrtnat}
\bibliography{main.bib}

%\pagebreak
% \input{appendix}

%TC:endignore
\end{document}